\documentclass[11pt]{article}
\usepackage[a4paper, total={6in, 8in}]{geometry}
\RequirePackage[OT1]{fontenc}
\RequirePackage{amsthm,amsmath, lscape, bm, multirow, graphicx}

\RequirePackage[numbers, super, sort]{natbib}
\RequirePackage[colorlinks,citecolor=blue,urlcolor=blue]{hyperref}

\usepackage{authblk}
\title{A hierarchical modelling approach to assess multi pollutant effects in time-series studies}
\author[1,*]{Marta Blangiardo}
\author[1]{Monica Pirani}
\author[2]{Lauren Kanapka}
\author[1]{Anna Hansell}
\author[3]{Gary Fuller}

\affil[*]{Corresponding author}
\affil[1]{MRC-PHE Centre for Environment and Health,
Department of Epidemiology and Biostatistics,
Imperial College,
St Mary's Campus,
Norfolk Place, London W2 1PG}
\affil[2]{Department of Mathematics,
Imperial College,
South Kensington Campus,
Exhibition Road, London SW7 2AZ}
\affil[3]{MRC-PHE Centre for Environment and Health,
Environmental Research Group,
King's College,
150 Stamford Street,
London SE1 9NH}

\date{}                     %% if you don't need date to appear
\begin{document}
  \maketitle

\begin{abstract}
When assessing the short term effect of air pollution on health outcomes, it is common practice to consider one pollutant at a time, due to their high correlation. Multi pollutant methods have been recently proposed, mainly consisting of collapsing the different pollutants into air quality indexes or clustering the pollutants and then evaluating the effect of each cluster on the health outcome. A major drawback of such approaches is that it is not possible to evaluate  the health impact of each pollutant. In this paper we propose the use of the Bayesian hierarchical framework  to deal with multi pollutant concentration in a two-component model: a pollutant model is specified to estimate the `true' concentration values for if your  each pollutant and then such concentration is linked to the health outcomes in a time series perspective. Through a simulation study we evaluate the model performance and we apply the modelling framework to investigate the effect of six pollutants on cardiovascular mortality in Greater London in 2011-2012.
\end{abstract}

%\begin{keyword}[class=MSC]
%\kwd[Primary ]{60K35}
%\kwd{60K35}
%\kwd[; secondary ]{60K35}
%\end{keyword}

%\begin{keyword}
%\kwd{Multi pollutant concentration}
%\kwd{Bayesian hierarchical model}
%\kwd{short-term health effects}
%\kwd{uncertainty propagation}
%\end{keyword}

%\jnlcitation{\cname{%
%\author{Blangiardo M.}, 
%\author{M. Pirani}, 
%\author{L. Kanapka}, 
%\author{A. Hansell}, and 
%\author{T. Woollings}} (\cyear{2016}), 
%\ctitle{A regime analysis of Atlantic winter jet variability applied to evaluate HadGEM3-GC2}, \cjournal{Q.J.R. Meteorol. Soc.}, \cvol{2017;00:1--6}.}

\section{Introduction}
Short-term air pollution studies aim at evaluating the association
between the day-to-day variation in ambient air pollution and the day-to-day variation
in a health outcome, such as mortality or hospital admissions. Typically this involves a time-series approach
using data from a particular geographical area that contains daily counts of mortality
or morbidity, pollution and meteorological measurements. To provide a few recent examples, Xie \textit{et al.} \citep{Xie:2014} used a data set from Beijing containing counts of daily hospital admissions and mortality from ischaemic heart disease (IHD), measurements of fine particulate matter
air pollution and potential confounding meteorological variables such as temperature or relative
humidity. Using a Poisson model, they reported a
significant association between the pollutant and IHD. Other studies using a time-series
approach \citep{Atkinson:2016}, \citep{Samoli:2016}, \citep{Dai:2014}
had mixed results in detecting a short-term association between various air pollutants
and a mortality outcome.

It is obvious that the air that we breath contains a number of different pollutants; however, due to the high correlation between these, the typical approach used in this field evaluates the health effect of one pollutant (or at most two pollutants) at a time. Despite this, recently there have been some attempts to move towards a multi pollutant approach that more realistically depicts the complexity of the exposure. For instance, Pirani \textit{et al.} \citep{Pirani:2015} proposed a Dirichlet
process (DP) mixture model to cluster the days by their exposure profiles. The model jointly estimates the covariate patterns and the health effect of each cluster. In addition the use of the DP allows the number of clusters to be determined
by the model and the data, guaranteeing extreme flexibility.

In a similar perspective Bobb \textit{et al.} \citep{Bobb:2015} proposed a Bayesian Kernel
Machine Regression (BKMR). The idea is to include the pollutants in the
model using a smooth function \textit{h} that is represented through a kernel function. The authors focused on Gaussian kernel as it outperformed linear and ridge regression kernels in simulation studies where \textit{h} has a complex functional form.

Similarly to the DP approach, the focus of BKMR is to correctly
identify the exposure-response relationship rather than to identify the effect of each individual pollutant on the outcome. To partially address this,
the authors extended their model to include a framework for variable selection, allowing the inclusion only of pollutants which have an impact on the response. However, such approach does not quantify this impact to determine exactly how much the pollutants are
affecting the response.

An alternative approach to deal with multiple pollutants consists of building a composite air quality index, that summarises their concentration; this is commonly carried out by most governments and used to inform people of the health risk posed by air quality. For example, in the UK the Daily Air Quality Index (DAQI) is based on the highest pollutant concentration out of the following regulated ones: sulphur dioxide, ozone, particulate matters, nitrogen dioxide. The concentration is then transformed into pre-determined bands (low, moderate, high, very high). Recently DAQI has been used to evaluate the effect of episodes of high air pollution concentration on respiratory conditions  \citep{Smith:2015}. %The technique has been also extended to account for geostatistical setting \citep{Lee:2011,Powell:2014}.

Finazzi \textit{et al.} \citep{Finazzi:2013} proposed a more sophisticated approach by using a hierarchical
model based on latent variables, that they called dynamic co-regionalization model. This
model aggregates the pollutant data over space and addresses problems such as missing
data and an unbalanced network, where not all pollutants are measured at every site.
The output of this model can then be used to calculate an index, for instance by taking the maximum, such as the
DAQI does. The authors applied their model to Scottish air pollution data and used the maximum to calculate a state-wide index over time.

Very recently Huang \textit{et al.} \citep{Huang:2017} proposed a two-stage spatio-temporal approach to evaluate the effects of two pollutants on respiratory hospital admissions in Scotland. In the first step they estimate annual concentrations from monitoring stations and output from numerical models and then feed forward the estimates and their uncertainty to the second step to assess the health effects. To include the two pollutants in the second stage, avoiding collinearity, they consider the first pollutant and the residual of the second after accounting for the first through a linear regression, making the approach difficult to be extended to more than two pollutants. 

Our paper is set in a similar perspective as we develop a two-components Bayesian hierarchical model that quantifies the health effect of multi pollutants. In the first component we account for measurement error in the observed air pollution measurements and for correlation among pollutants; based on this we estimate the corresponding latent `true' concentration values. However, our paper novelty lays on its fully Bayesian framework, as the two components are jointly estimated so that uncertainty from the concentration estimates can feed forward into the health effect estimates; at the same time information from the outcome can feedback to the air pollution estimates. We do rely on the joint estimation process, on the hierarchical nature of the model and on informative priors on the health effect parameters to overcome the collinearity among the pollutants, making the framework extendible to any number of pollutants, hence able to disentangle synergic or antagonistic effects of pollutants which would be not detectable in the common single-pollutant modelling framework. The developed approach is used to evaluate the effect of five pollutants (carbon monoxide - CO, nitrogen dioxide - NO$_2$, ozone - O$_3$, sulphur dioxide - SO$_2$ and fine particulate matter, smaller in size than 2.5 $mg/m^{3}$ - PM$_{2.5}$) and particle number concentration - PCNT on daily cardiovascular mortality in Greater London for 2011-2012. 

The remainder of the paper is structured as follows: section \ref{Sec:MatMet} presents the data and the model, section \ref{Sec:Sim} introduces the simulation study, while in section \ref{Sec:Results} we present the results of our analysis and section \ref{Sec:Discussion} covers areas of discussion and concluding remarks.

\section{Material and Methods}\label{Sec:MatMet}
\subsection{Data Description}\label{Sec:Data}
Daily measurements of CO, NO$_2$, O$_3$, SO$_2$, PM$_{2.5}$ and PCNT were obtained from a monitoring site in North Kensington, London (UK) over the period 1 January 2011 to 31 December 2012.
The London North Kensington site (lat 5131015.78000 N, long 012048.57100 W) is part of both the London Air Quality Network and the national Automatic Urban and Rural Network and is owned and part-funded by the Royal Borough of Kensington and Chelsea. The facility is located within a self-contained cabin on a school ground in a mainly residential area. It  has been used in previous time-series studies on air pollution health effects \citep{Atkinson:2010},\citep{Pirani:2015},\citep{Atkinson:2016}. The same monitoring site has also been used extensively as a background measurement site for source apportionment \citep{Beddows:2015} and also to track the outcome of policies to improve London air pollution \citep{Font:2016}.

CO, NO$_2$, O$_3$ and SO$_2$ were measured using CEN mandated methods eg EN 14211 for NO$_2$. Fortnightly calibrations enabled the traceability of measurements to national meteorological standards. PM$_{2.5}$ were measured by TEOM-FDMS (Tapered Element Oscillating Microbalance - Filter Dynamics Measurement System) which is considered equivalent to the EU reference method. Particle number concentration was measured by condensation particle counter (TSI 3022).

We focus on these five pollutants as they are already regulated in ambient air; as a result, they are well monitored, have documented associations to health outcomes \citep{REVIHAPP} and have been showed to need National Ambient Air Quality Standards \citep{Owens:2017}. In addition several papers have focused on one or more of these: for instance Mills \textit{et al.} \citep{Mills:2016} presented a systematic review of the effects on NO$_2$ where particulate matter is also controlled for. Besides the five pollutants, we also investigate the effect of PCNT, as this metric was previously associated  to adverse short-term health outcome in London \citep{Atkinson:2010}.

As a health outcome we consider the daily count of mortality due to cardiovascular diseases (CVD) over the same period obtained from the UK Office of National Statistics and available through the Small Area Health Statistics Unit (SAHSU). These cardiovascular causes were derived from the International Statistical Classification of Diseases, 10th Revision (ICD-10, Chapter I).

To adjust for potential confounding effect of weather variables, we use daily average temperature and relative humidity obtained from a meteorological station close to the North Kensington monitoring site. Table \ref{summary} provides descriptive statistics of the variables considered in the analysis.
\begin{table}
\caption{Descriptive statistics of the variables included in the study\label{summary}}
\centering
\begin{tabular}{lccccccc}
\hline
& Number & \multicolumn{5}{c}{Percentiles} & \\
& of Days & 10th & 25th & 50th & 75th & 90th  & IQR\\
\hline
Mortality & 731 & 28 & 32 & 37 & 42 & 47 & 10\\
\emph{Meteorological data}: & & & & & & & \\
Temperature ($^\circ C$) & 731 & 5.1 & 8.0 & 11.7 & 15.5 & 18.1 & 7.4\\
Relative Humidity (\%) & 731 & 61.6 & 69.6& 78.0& 84.2 &88.5 & 14.5\\
\emph{Pollutants}: & & & & & & & \\
$\quad$ CO ($mg/m^3$) & 715 & 0.1 &0.2 &0.2 &0.3& 0.4& 0.1\\
$\quad$ NO$_2$ ($\mu g/m^3$) & 706 & 18.2 & 23.2& 33.3 & 46.9 & 57.9 & 23.6\\
$\quad$ O$_3$ ($\mu g/m^3$) & 695 & 11.4 & 24.3 & 39.1 & 51.1 & 64.9 & 26.8\\
$\quad$ SO$_2$ ($\mu g/m^3$) & 717 & 0.0 & 0.4 & 1.8 & 2.6 & 3.6 & 2.2\\
$\quad$ PM$_{2.5}$ ($\mu g/m^3$) & 730 &  5.0  & 6.0  & 9.0 & 14.0  &25.0 & 8.0\\
$\quad$ PCNT ($p/mm^3$)& 636& 7.8 & 9.7 &12.1& 14.9 & 17.9 & 5.2 \\
\hline
\end{tabular}
\end{table}

\subsection{Model Specification}\label{Sec:Model}
Our modelling framework consists of two components jointly estimated: a pollutant model and a health model, which we describe in details in this section.

\subsubsection{Pollutant model}
We start specifying $Y_{pt}$ as the measured concentration level of pollutant $p$ ($p=1,...,P=6$) on day $t$ ($t=1,...,T=731$) from the monitoring site. As different pollutants are typically characterised by different scales we recommend standardisation to make them comparable. As $Y_{pt}$ is a continuous variable it is reasonable to assume the following Normal distribution:
\begin{equation}
Y_{pt}\sim N(\mu_{pt}, \sigma^2_p)
\label{observationEquation}
\end{equation}
where $\sigma^2_p$ is the measurement error variance, which is specific for each pollutant. On $\mu_{pt}$ a linear model is specified as follows:
\begin{equation}\label{Eq:pol}
\mu_{pt} = \gamma_{0p} + \gamma_{1p}\text{X}_{\text{temp,$t$}} +\gamma_{2p}\text{X}_{\text{temp,$t$}}^2 + \gamma_{3p}\text{X}_{\text{rhum,$t$}} + \gamma_{4p}\text{X}_{\text{rhum,$t$}}^2 + \theta_{pt}
\end{equation}
where $\gamma_{0p}$ is the pollutant specific intercept, while $\boldsymbol{\gamma}_{p}$ are the regression coefficients linking the time-dependent covariates $\bm{X}_{t}$ to the pollutant levels; as descriptive plots suggest the presence of a non-linear relationship between the covariates and the pollutant concentration levels (see Figure 1 in Supplementary material), we include a linear and quadratic effect of both temperature and relative humidity. In (\ref{Eq:pol}) $\{\theta_{1t}, \ldots, \theta_{Pt}\}$ account for the residual temporal effects and for the correlation among pollutants;  they are modelled following a multivariate Normal specification with an autoregressive structure, following Shaddick \textit{et al.} \citep{Shaddick:2002}:
\begin{equation}\label{Eq:theta}
(\theta_{1t},...,\theta_{Pt})^{'} \sim \text{MVN}\left((\theta_{1,t-\ell},...,\theta_{P,t-\ell})^{'}, \Sigma_P\right)
\end{equation}
where $t-\ell$ provides the temporal lag of $\ell$ days for the $t$-th day.
For each pollutant, the concentration at time $t$ depends on the values at time $t-\ell$, while the diagonal of the covariance matrix of the errors $\Sigma_P$ allows each pollutant to have a different amount of temporal dependence, with larger values indicating a smaller dependence. The off-diagonals represent the temporal dependence between the pollutants, allowing the model to incorporate and maintain the correlation structure in the estimation of the `true' pollutant levels.

Note that this specification has the added benefit of providing a natural way to deal with missing data in the pollutant concentration. As seen in Table \ref{summary}, there are some days where the concentration is not available for one or more pollutants; the model could impute directly the concentration based on the correlation with the other pollutants and on the temporal dependency.
\subsubsection{Health Model}
The second component of the model links the `true' latent value of the pollutant concentrations $\mu_{pt}$ with the counts of the health outcome within a time-series epidemiological framework. Let O$_{t}$ be the observed number of CVD deaths for the day $t$, we specify a Poisson distribution as:
\begin{equation}\label{Eq:O}
\text{O}_t \sim \text{Poisson}(\lambda_{t} \text{E}_t)
\end{equation}
where E$_t$ represents the expected number of CVD deaths. Following Pirani \textit{et al.} \citep{Pirani:2015}, we take it to be the average mortality over the whole period, hence E$_t$=E. Then $\lambda_{t}$ represents the relative risk of CVD death on day $t$ compared to the average. In a previous analysis performed on the same data set, Atkinson \textit{et al.} \citep{Atkinson:2016} considered the association between CVD and 1-day lagged pollutant concentrations, thus we coherently adopt the same exposure window setting $\ell=1$.  We therefore specify a regression model on the log link transformed $\lambda_{t}$:
\begin{equation}\label{Eq:Mort}
\log(\lambda_{t})= \beta_0 + \sum_p \beta_p \mu_{p(t-1)} + \sum_i s(Z_{ti}, \psi_i) + \delta_{I_t} + \epsilon_t
\end{equation}
so that $\exp(\beta_p)$ is the multiplicative change in relative risk of CVD death for a unit increase in the pollution concentration obtained from (\ref{Eq:pol}). To be able to interpret the health effects on the correct scale, we back transform the pollutant concentration $\mu_{pt}$ estimated from  \eqref{Eq:pol} to the  original scale.
In \eqref{Eq:Mort} $s(\cdot, \psi_i)$ denote smooth functions of daily average temperature and relative humidity, as well as of calendar time to account for any residual seasonality and long-term trends. These confounding factors are included in the model through flexible nonparametric penalised spline functions \citep{Ruppert:2003}. In particular, we consider a mixed model framework and following Crainiceanu \textit{et al.} \citep{Crainiceanu:2005} we specify a low-rank thin plate spline basis over other options, which tends to show a smaller posterior correlation between parameters. By letting $Z_{ti}$ be the $i$-th confounder on day $t$, we have the following spline representation:
\begin{equation}\label{Eq:Spline}
s(Z_{ti}, \psi_{i})= \alpha_{i} Z_{ti} + \sum_{k=1}^{K_{i}} b_{ki} | Z_{ti}-\kappa_{ki}|^3
\end{equation}
where $\psi_i=(\alpha_{i}, b_{1i},..., b_{Ki})^{'}$ are the regression coefficients, $ Z_{ti} -\kappa_{ki}$ are the set of basis functions of the cubic spline and $K_{i}$ is the number of knots for confounder $i$, with knot locations $\kappa_{1i}<\kappa_{2i}<...<\kappa_{Ki}$. Based on Atkinson \textit{et al.} \citep{Atkinson:2016} we select 3 knots for temperature and relative humidity and 6 for time.
Additionally, to account for any holiday effect, we include in the model the linear term $I_t$ which classifies the days according to workday or weekend/holiday. Finally to account for overdispersion, that is typically present when a Poisson distribution is assumed on the data, we include an additional random effect $\epsilon_t \sim N(0,\sigma_{\epsilon}^2)$.

\subsection{Prior Specification}
The last step in the model specification consists of the choice of prior distributions. Minimally informative Normal distributions are specified on all the regression coefficients $\gamma_0$, $\boldsymbol{\gamma}_{p}$, $\beta_0$, $\delta$ and $\boldsymbol{\alpha}$,  centered on 0 and with a variance equal to 10$^3$. Given the high correlation present among the pollutants, we take advantage of an informative prior on $\beta_p$; we choose a $N(0,0.1)$ covering a range of values on the relative rate scale from 0.82 to 1.22, which is plausible with what has been seen in the literature on cardio-respiratory diseases.

On the standard deviation for the measurement error $\sigma_p$ and for the random effect $\sigma_{\epsilon}$ a Uniform prior is specified ranging between 0 and 100, to ensure minimal information.

The covariance matrix $\Sigma_P$ is given a $P$-dimensional inverse Wishart prior, $IW(D,d)$, where $D$ is a symmetric and positive-definite scale matrix and $d$ is the degrees of freedom parameter. We follow the specification presented in Lunn \textit{et al.} \citep{bugs:2012} and to ensure the weakest information we fix $d=P$; as the prior mean for the inverse Wishart is $d^{-1}D$, $D$ is chosen to be $d$ times the prior estimate of the correlation matrix.

We penalise the random coefficients associated to the basis functions, $b_{ki}$, shrinking them towards zero to avoid over-fitting. We assume a Normal prior distribution for those coefficients, with mean 0 and unknown precision specific for each confounder $\sigma^{-2}_{b_{i}}$. This latter parameter controls the amount of smoothness and is supplemented with a $\text{Gamma}(a,b)$ prior distribution, where $a=1; b=0.001$.

\subsection{Implementation and Sensitivity Analysis}
The model is run using a MCMC simulative framework in R; we discarded the first 50,000 iterations of the MCMC and retained the following 10,000 to estimate the posterior distribution of the parameters. We considered two chains and checked for convergence of the parameters visually (see Figures 2-4 in Supplementary Material) and analytically (evaluating the MC error below 5\% of the standard deviation of the posterior estimates as well as the Gelman-Rubin diagnostic tool).

It is important to stress that this is the first paper to consider jointly the pollutant and the health components; this results in uncertainty on $\mu_{pt}$ affecting the estimates of the relative risks $\beta_1,\ldots, \beta_P$, while at the same time the information from the outcome is fed backwards into the latent concentration values. This is a crucial point as in this way the correlation between the pollutants is naturally accounted for through the hierarchical structure and through the input from the outcome. 

To evaluate the robustness of our modelling framework we changed the prior specification of all the parameters where a no informative prior was assumed. In particular on the regression coefficients we specified a Normal distribution centered on zero and with a variance equal to 10$^6$, while the measurement error and random effect variance were set to Inverse Gammas with parameters 1 and 0.001; finally we put a Gamma with parameters $a=b=0.001$ on the precisions of the random coefficients associated to the basis functions.

A key aspect in air pollution time-series health studies is represented by the inclusion of the smoothing functions for the time-varying confounding factors. Here, we need to adequately control for their potential non-linear confounding effect while retaining sufficient information for estimating the exposure effects. To perform model checking on the knots we also ran the model with 14 knots on time (7 for each year, \citep{Dominici:2000}), and 9 on temperature and humidity, which we think is large enough to account for a high degree of non linearity, while at the same time not leading to oversmoothing. As model selection tool we used the Deviance Information Criterion (DIC, Spiegelhalter \textit{et al.} \citep{Spiegelhalter:2002}), one of the suggested methods\citep{Peng:2006} to choose the  degree of smoothness for time-series studies of air pollution and mortality.

\section{Simulation Study}\label{Sec:Sim}
We carried out a simulation study to evaluate if the proposed modelling framework is able to estimate the relative risk of highly correlated pollutants on a health outcomes.
\subsection{Simulation set-up}
We simulated mortality and air pollution concentration for 2000 days and considered 6 pollutants. For the sake of simplicity we did not include any confounder factor (e.g. meteorology) in the pollutant or health components of the model. We fixed the correlation among the pollutants to be equivalent to that observed on the time-series data from Greater London:
\begin{equation}\label{Eq:Corr}
\multirow{6}{*}{P$_P=  $} \left[{\begin{array}{cccccc}
1 & 0.737 & -0.535 & 0.442 & 0.515 & 0.630\\
  & 1 & -0.606 &  0.510 & 0.730 & 0.659\\
 & & 1 & -0.260 & -0.394 & -0.396\\
 & & & 1 & 0.390 & 0.490\\
 & & & & 1 & 0.420\\
  & & & &  & 1\\
 \end{array}}\right].
\end{equation}
The following steps were used to simulate the data on concentration and outcome:
\begin{enumerate}
\item Using the above correlation matrix we generated the true pollutant levels assuming an autoregressive structure of order 1, as specified in (\ref{Eq:theta}), $\mu_{pt} \sim N(\mu_{p(t-1)}, \text{P}_P)$. This represents the gold standard exposure.
\item At the same time we also simulated the measured concentration for the six pollutants, which we assumed centered on the true latent exposure, but with a measurement error variance equal to 0.1 ($Y_{pt} \sim N(\mu_{pt}, 0.1)$).
\item We then simulated the daily number of events for a health outcome using a Poisson distribution, where the mean $\lambda_t$ is specified as
\begin{eqnarray*}
\log(\lambda_t) = 1 + 0.2 \mu_{t1} + 0.2 \mu_{t2} - 0.2 \mu_{t3} +  0 \mu_{t4} + 0 \mu_{t5} + 0 \mu_{t6}
\end{eqnarray*}

so that we are able to assess if the model can capture true effects as well as the lack thereof.
 
\item We repeated the process 100 times.
\end{enumerate}

We ran our modelling framework, hereafter called ``hierarchical two-component model'' (\textit{H2Mjoint}), and compared it with a standard Poisson model, here named as ``measurement error model'' (\textit{ME}), where the true concentration is replaced by the measured one:
\begin{eqnarray}\label{Eq:SP}
O_t &\sim& \text{Poisson}(\lambda_{t}E_t)\nonumber\\
\log(\lambda_{t})&=& \beta_0 + \sum_p \beta_p Y_{p(t-1)} + \sum_i s(Z_{ti}, \psi_i) + \delta_{I_t} +\epsilon_t
\end{eqnarray}

\begin{figure}[!h]
%\centering
\includegraphics{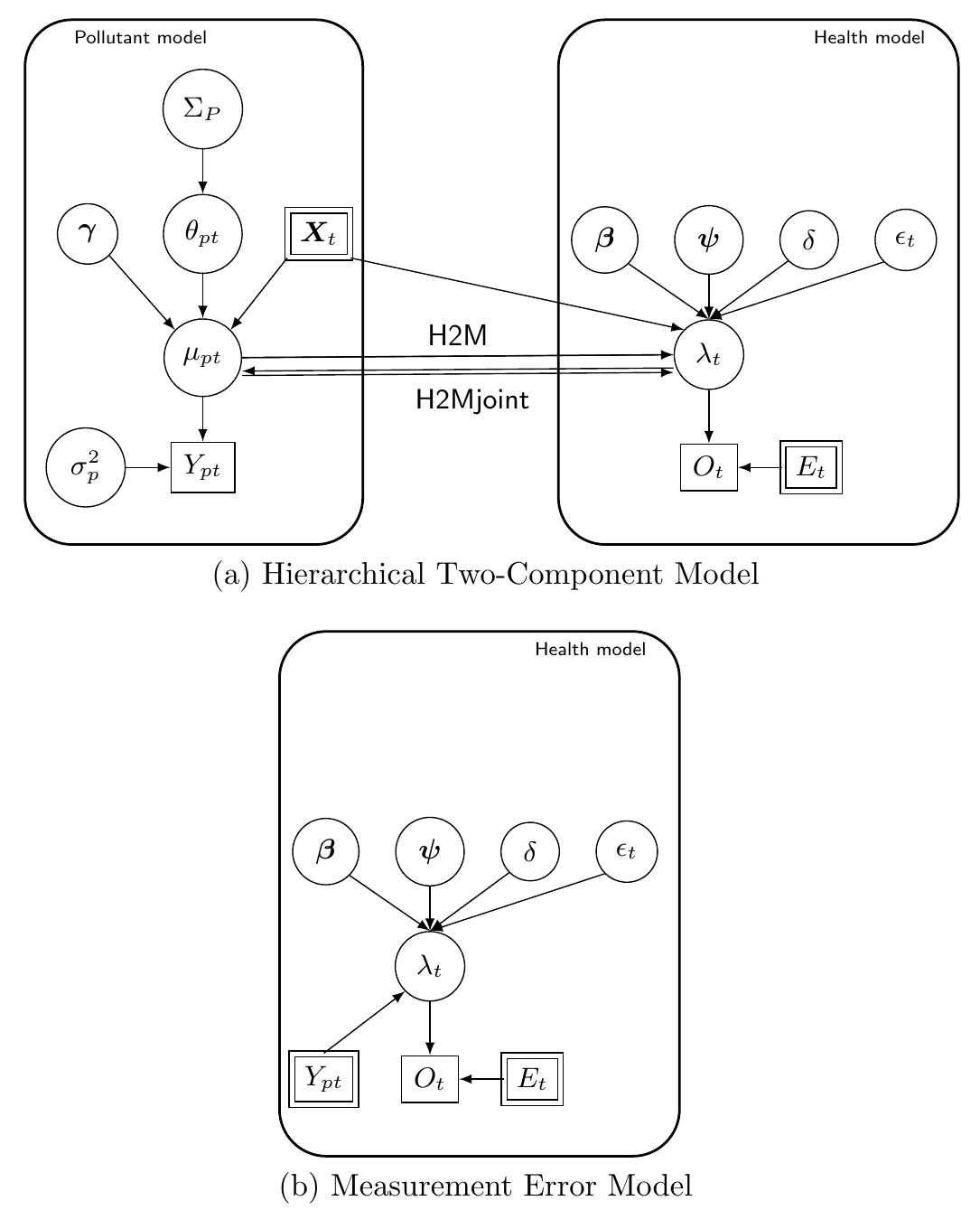}
\caption{\textbf {Graphical representation of the modelling frameworks:} (a) shows the proposed two component model: the left hand side represents the pollutant component, while the right hand side the health component. The latent concentration for each pollutant and day, $\mu_{pt}$, obtained from the pollutant component enters the health model as predictor. The specification of the link between $\mu_{pt}$ and $\lambda_t$ makes the difference between H2M and H2Mjoint. In the former the uncertainty from $\mu_{pt}$ goes forward into the health model, but there is no feedback from $O_t$; in the latter the uncertainty goes forward, while at the same time information from the mortality count $O_t$ can influence back $\mu_{pt}$. (b) shows the ME model: the pollutant component is not there and the measured pollutant concentration $Y_{pt}$ is now directly linked to $\lambda_t$. For both (a) and (b) the circles denote latent random variables, while the rectangles are observed quantities; single rectangles are random variables, while double rectangles enter the model as data and are not characterised by a probability distribution.}\label{Fig:DAG}
\end{figure}

We used this as benchmark, given that it is the model commonly specified in epidemiological studies to study short-term health effects of air pollution. The graphical representation of ME is presented in Figure \ref{Fig:DAG}(b) and shows a direct link between $Y_{pt}$ and $\lambda_{t}$.

The H2Mjoint model jointly estimates the pollutant concentrations and their health effects through a fully specified Bayesian framework. As an additional comparison we specified an alternative model where the two components are fitted separately, called \textit{H2M}: for this model first \eqref{observationEquation}-\eqref{Eq:theta} are run and the posterior distribution for the pollution concentration is estimated. The distribution is then fed forward into the health component (\eqref{Eq:O}-\eqref{Eq:Spline}), so that the air pollution effects on health account for the uncertainty which derives from their estimated concentration. At the same time in H2M, the feedback from the health outcome is not allowed to influence the air pollutant concentration estimates. H2Mjoint and H2M are both represented in Figure \ref{Fig:DAG}(a): the former has two links between $\mu_{pt}$ and $\lambda_t$, going each in one direction (uncertainty feeding forward and backwards), while for the latter there is only one link and the arrow only points from $\mu_{pt}$ to $\lambda_t$ as no feedback is allowed.
The model comparison is carried out in terms of bias, root mean square error (RMSE), 95\% credible interval (CI) coverage and 95\% CI width.

\subsection{Simulation results}
\begin{figure}[!h]\begin{center}
\centering\includegraphics[scale=0.55]{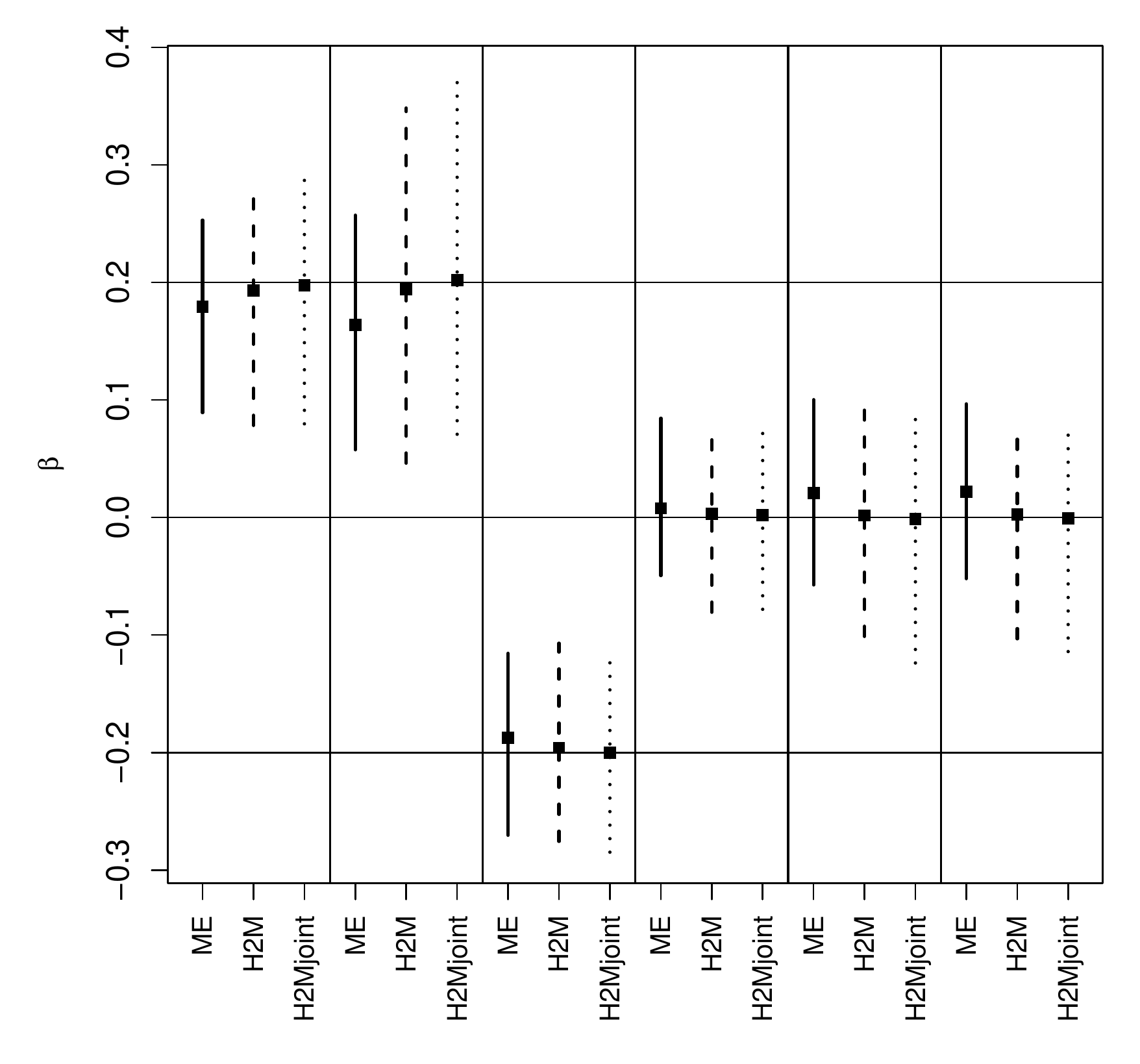}
\caption{\textbf{95\% posterior credible intervals for $\boldsymbol{\beta}$ under ME, H2M and H2Mjoint.} The H2Ms show smaller levels of uncertainty, as this influence the coefficients from the pollutant estimates as well as from the health model itself. At the same time the ME model shows a larger bias in the estimates, due to the measurement error, while H2Mjoint model show a median estimate virtually equal to the true values, showing how the fedeback from the outcome can play a role in reducing the corresponding bias.\label{Fig:CIsim}}\end{center}
\end{figure}

Table \ref{Tab:SimRes} presents the results of the simulation study in terms of the indexes above. It is clear that the hierarchical two-component model framework (H2Mjoint / H2M) outperforms the model that uses measured air pollution concentration (\textit{ME}, as in  \eqref{Eq:SP}); across the six pollutants the bias is reduced by 3 to 11 fold and the coverage of the CI95\% is always above 90\% (compared to 53-77\% for the \textit{ME} model). In terms of precision, the RMSE is generally smaller for the Bayesian model, indicating better accuracy in the estimates, while the width of the confidence interval is larger, which can be explained by the additional uncertainty included in the concentration estimates and that feeds forward into the health component. This can also be seen in the 95\% CI plot for the $\boldsymbol{\beta}$ coefficients (Figure \ref{Fig:CIsim}). Comparing H2Mjoint with H2M shows that there is an advantage in allowing for the joint specification of the two model components: in H2Mjoint the bias is smaller, which is clearer when the true effects are different from 0 ($\beta_1$-$\beta_3$); at the same time there is no increase in the estimate uncertainty, as the widths of the 95\% credible intervals do not change substantially. 
%
%\begin{landscape}
\begin{table}[h!]	
\begin{center}\caption{Results of the simulation study: the table shows the bias, root mean square error (RMSE), 95\% credible intervals (CI) width and coverage for the ME, H2M and H2Mjoint. The bias and RMSE are substantially reduced for all the 6 pollutant coefficients using H2Mjoint / H2M. Coverage improves and at the same time width of the 95\% credible interval increases, suggesting that the uncertainty is larger for the hierarchical two-component modelling framework, as expected, given that this comes also from the pollutant component. The comparison of H2Mjoint with H2M shows how the influence of the outcome helps reduce the bias, while at the same time the uncertainty does not increase.\label{Tab:SimRes}}
\centering
\begin{tabular}{ccccccc}
\hline
& \multicolumn{3}{c}{Bias}	& \multicolumn{3}{c}{RMSE}\\
\hline
	&	ME	&	H2M	&	H2Mjoint	&	ME	&	H2M	&	H2Mjoint\\
$\beta_1$	&	-0.021	&	-0.007	&	-0.002	&	0.003	&	0.002	&	0.002	\\
$\beta_2$	&	-0.036	&	-0.006	&	0.002	&	0.004	&	0.005	&	0.004	\\
$\beta_3$	&	0.013	&	0.004	&	0.000	&	0.003	&	0.004	&	0.002	\\
$\beta_4$	&	0.008	&	0.003	&	0.002	&	0.001	&	0.002	&	0.001	\\
$\beta_5$	&	0.021	&	0.002	&	-0.001	&	0.002	&	0.002	&	0.002	\\
$\beta_6$	&	0.022	&	0.002	&	-0.001	&	0.002	&	0.002	&	0.002	\\
\hline
& \multicolumn{3}{c}{95\% CI width}	& \multicolumn{3}{c}{95\% CI coverage}\\
\hline
	&		ME	&	H2M	&	H2Mjoint	&	ME			&	H2M	&	H2Mjoint	\\
$\beta_1$	&	0.16	&	0.20	&	0.20	&	65	&	92	&	93	\\
$\beta_2$	&	0.20	&	0.30	&	0.30	&	53	&	97	&	97	\\
$\beta_3$	&	0.16	&	0.17	&	0.16	&	71	&	92	&	97	\\
$\beta_4$	&	0.13	&	0.16	&	0.16	&	77	&	98	&	99	\\
$\beta_5$	&	0.16	&	0.19	&	0.22	&	61	&	95	&	94	\\
$\beta_6$	&	0.15	&	0.18	&	0.19	&	65	&	97	&	99	\\
\hline
\end{tabular}\end{center}
\end{table}
%\end{landscape}

%
\section{Real application results}\label{Sec:Results}
We focused on H2Mjoint to evaluate the effects of the six pollutants (CO, NO$_2$, O$_3$,  SO$_2$, PM$_{2.5}$ and PCNT) on daily CVD mortality in Greater London for 2011-2012, as the simulation showed that allowing for the feedback from the outcome leads to an improvement in the estimates in terms of bias. In addition to the multi pollutant H2Mjoint, we ran single pollutant models as a comparison, given that this is the typical approach in the field. The DIC was smaller for the model with 6 knots on time and 3 on temperature and humidity (9257), thus we present the results of this model specification (which was also the one used in Atkinson \textit{et al.}\citep{Atkinson:2016}). For the models with increased number of knots and different prior the results are presented in Table 1 and 2 of Supplementary material, together with their DIC. 

The model is able to reproduce the temporal pattern seen in the data, as shown in the residuals (Figure 5 of Supplementary Material), that are scattered around 0 for all the six pollutants as well as for the mortality counts. 

As the pollutants are on different scales, to make their effects comparable, in Table \ref{Tab:Res} we present the results in terms of percent increase for a interquartile range (IQR) change in air pollution concentrations, defined as:
\[\% \text{increase} = (e^{\beta_p \times IQR} - 1) \times 100\%\]

\begin{table}[hbt]
\caption{Posterior mean and 95\% credible interval of the percent increase in mortality for an IQR change in pollutant concentration: (left) multi pollutant H2Mjoint model; (centre) single pollutant H2Mjoint model; (right) single pollutant frequentist model (Atkinson \textit{et al.}, 2016). Note that all the pollutants are measured in $\mu g/m^3$ except for PCNT which is measured in $p/mm^3$.
\label{Tab:Res}}
\centering
\begin{tabular}{lrrlrlrl}
\hline
\multicolumn{1}{c}{} & \multicolumn{1}{c}{} & \multicolumn{2}{c}{Multi Pollutants} & \multicolumn{2}{c}{Single Pollutants} & \multicolumn{2}{c}{Single Pollutants}\\
\multicolumn{1}{c}{} & \multicolumn{1}{c}{} & \multicolumn{2}{c}{(H2Mjoint)} & \multicolumn{2}{c}{(H2Mjoint)} & \multicolumn{2}{c}{(Atkinson et al., 2016)}\\
\multicolumn{1}{c}{\textbf{Pollutant}} &   \multicolumn{1}{c}{\textbf{IQR}} & \multicolumn{2}{c}{\textbf{\% Increase}} & \multicolumn{2}{c}{\textbf{\% Increase}} & \multicolumn{2}{c}{\textbf{\% Increase}}\\
 &    & \multicolumn{2}{c}{\textbf{(95\% CI)}} & \multicolumn{2}{c}{\textbf{(95\% CI)}} & \multicolumn{2}{c}{\textbf{(95\% CI)}}\\
\hline
CO& 0.10 &  -1.67 &(-4.72, \hspace{2pt}1.65) &  -1.59 &(-3.89, 0.84) & * &  *\\
NO$_2$&23.65  & 9.40 &(3.06, 16.03) &  -0.25 &(-2.90, 2.43) & -1.69 & (-3.97,  0.64)\\
O$_3$&26.85&  3.46 &(0.18, \hspace{4pt}6.71) & 2.61 &(0.02, 5.32) & 3.31& ( 0.83,  5.84)\\
SO$_2$ & 2.20  & -1.94 & (-6.59, \hspace{2pt}2.80)  & -1.13 & (-4.96, 3.15) & -2.33 & (-4.18, -0.45)\\
PM$_{2.5}$&8.00& -1.24 &(-3.45, \hspace{2pt}0.92) & -0.79 &(-2.06, 0.47) & -0.9 & (-2.09, 0.25)\\
PCNT &5.18& -2.89 &(-6.36, \hspace{2pt}1.05) &  -0.31 &(-3.56, 3.35) & * & *\\
\hline
\multicolumn{8}{l}{$^*$ CO and PCNT were not analysed in Atkinson \textit{et al.}, 2016.}
\end{tabular}
\end{table}

Out of the six metrics NO$_2$ and O$_3$ shows an increased risk of CVD mortality, with credible intervals entirely above 0, suggesting strong evidence of an effect. For the remaining pollutants the point estimates of percent change are slightly below 0, but there is a high degree of uncertainty on the results and the credible intervals include 0. There is high correlation between measured and latent pollutant concentration and, as expected, the latter is slightly less extreme due to shrinkage intrinsic in the modelling framework (see figure 6 in Supplementary material for a plot comparing the posterior mean of $\mu_{pt}$ with the measured concentration $Y_{pt}$ for the six pollution metrics). In addition, the measurement error variance on the standardised metrics is presented on Table \ref{Tab:Var} and shows the lowest values for NO$_2$ and O$_3$, both around 0.04, while it is 0.08 for PM$_{2.5}$ and it increases between 0.14 to 0.45 for the remaining pollutants. This suggests that the model is able to account almost entirely for the variability of NO$_2$, O$_3$ and PM$_{2.5}$, while it would potentially point towards some residual confounding for SO$_2$, CO and PCNT.

By contrast the time-series single pollutant model shows a positive posterior mean with a credible interval above zero only for O$_3$. For the remaining pollutants the point estimates are negative and their intervals cross 0, pointing towards lack of substantive evidence of an effect. Measurement error variances are larger, with a posterior mean spanning from 0.11 to 0.66, suggesting that the multi pollutant model borrow strength across pollutants to improve the accuracy of the concentration estimates. 

The single pollutant results are in line with Atkinson \textit{et al.}\citep{Atkinson:2016}, who analysed the same period for Greater London and are reported on the right hand side of Table \ref{Tab:Res}. There is slightly more uncertainty in the H2Mjoint framework, as expected, as the pollutant component contributes to it. This translates into point estimates which are generally closer to zero and wider credible intervals; it is particularly interesting to note how accounting for uncertainty shift SO$_2$ estimates towards zero, so that the protective effect seen in Atkinson \textit{et al.}\citep{Atkinson:2016} disappears.

\begin{table}[hbt]
\caption{Posterior Mean and 95\% credible interval for the measurement error variance $\sigma^2_v$ for the H2Mjoint framework: multi pollutant model (left) and single pollutant model (right). Note that all the pollutants are measured in $\mu g/m^3$ except for PCNT which is measured in $p/mm^3$. \label{Tab:Var}}
\centering\begin{tabular}{lcccc}
\hline
\multicolumn{1}{c}{} & \multicolumn{2}{c}{Multi Pollutant Model} & \multicolumn{2}{c}{Single Pollutant Model}\\
%\multicolumn{1}{c}{} & \multicolumn{2}{c}{(H2Mjoint)} & \multicolumn{2}{c}{(H2Mjoint)}\\
\multicolumn{1}{c}{\textbf{Pollutant}} & \multicolumn{2}{c}{\textbf{Posterior Mean (95\% CI)}} & \multicolumn{2}{c}{\textbf{Posterior Mean (95\% CI)}}\\
\hline
CO&   0.18 &(0.15, \hspace{2pt}0.22) &  0.44 &(0.36, 0.52)\\
NO$_2$  & 0.03 &(0.01, 0.05) &  0.20 &(0.17, 0.27)\\
O$_3$&  0.04 &(0.02, \hspace{4pt}0.06) & 0.16 &(0.12, 0.21)\\
SO$_2$ & 0.45 & (0.29, \hspace{2pt}0.51)  & 0.66 & (0.55, 0.79)\\
PM$_{2.5}$& 0.08 &(0.04, \hspace{2pt}0.12) & 0.11 &(0.04, 0.18)\\
PCNT & 0.14 &(0.10, \hspace{2pt}0.17) &  0.59 &(0.49, 0.68)\\
\hline
\end{tabular}
\end{table}

\section{Discussion}\label{Sec:Discussion}

In this paper we proposed a fully Bayesian hierarchical model to assess the health effect of multi pollutant concentrations in a time-series perspective, allowing for the integration of uncertainty on the exposure and health components. We deal with the common issue of multi-collinearity among pollutants as the joint hierarchical specification allows to i) directly estimate and incorporate the correlation when modelling the `true` latent concentration from the measured ones at the monitoring site; ii) incorporate such correlation in the link between concentration and health. The use of a hierarchical model has been shown to provide stable estimates \citep{MacLehose:2007} and it allows to specify an informative prior, which acts as a constraint on the parameter estimates, helping deal with the potential colinearity among the pollutants. In addition, a Bayesian approach naturally accounts for missing data in the estimation process; this means that we were able to use the entire 731 days of the time-series, while other analyses \citep{Atkinson:2016} were based on less data points as the days with missing pollutant concentrations were removed. In our case, as we are considering six pollutants at the same time, this would mean removing 125 days as one or more pollutants did not have concentration value recorded. 

Note that the meteorological covariates (temperature and humidity) are included in the exposure component as well as in the health component of the model. This was done according to the approach proposed by Cefalu \textit{et al.} \citep{Cefalu:2014} who discussed how the covariates included in the exposure prediction model need to be included as confounders in the epidemiological model to avoid biased results.

A key characteristic of our modelling framework is that it involves a joint specification of the two components (pollution model and health model). In this way all the uncertainty is accounted for in the estimation process, differently from  classical two-stage models, where the pollution concentration estimates are considered without the associated uncertainty to evaluate their health effects. At the same time, in the simulation study, we found that the feedback from the outcome provides additional information to estimate the health effects when these are truly different from 0, while at the same time is not introducing bias in any direction when the true health effects are null.

Running our model on simulated data we found that the proposed framework caters well for the measurement error intrinsic in the observed concentrations and is able to estimate the health effects more accurately than the model which considers the observed concentrations as exposure in the health model. 

At the same time, on the real data application, we showed that our multi pollutant model is able to capture the short-term harmful effect of a possible synergic mechanism between NO$_2$ and O$_3$. After adjustment for airborne particles and other regulated gases (i.e. CO and SO$_2$), we found a positive association between a mixture of these two oxidant gases and cardiovascular mortality. The result is in line with Williams \textit{et al.}\citep{Williams:2014}, which considered a two-pollutant model of NO$_2$ and O$_3$, but for daily counts of all cause mortality for Greater London in 2000-2005, hence characterised by more power due to the longer period and larger numbers. The plausibility of the reported associations is also consistent with atmospheric chemistry findings \citep{Gamon:2014} and toxicological results \citep{Drechsler-Parks:1995}. Williams \textit{et al.}\citep{Williams:2014} have looked into combining the two pollutants, due to their high correlation and their complex chemistry; for instance NO$_2$ is a precursor of O$_3$, but also scavenging it, which explains why in the centre of cities the level of NO$_2$ is higher and the level of O$_3$ are lower. However a clear drawback  would be not to be able to disentangle the effects of the two pollutants, which might act on health through different mechanisms. For this reason we think that our approach is beneficial here, as it has shown the ability to identify and quantify the magnitude of the short-term health effect of the simultaneous exposure to multiple air pollutants, that it is not detectable using a traditional single pollutant model \citep{mauderly2009there}. Therefore, from an air quality management perspective we believe that a multi pollutant approach, such as the one proposed in this study, has a potential for suggesting effective control strategies to reduce adverse effects on human health, since it is able to provide insights on the complex trade-offs between different ambient pollutants.

In this paper we showed how the Bayesian hierarchical modelling framework is advantageous for dealing with multi pollutant concentrations in a purely time-series perspective. A natural extension will consists of increasing the number of measurement sites, moving to a spatio-temporal model. This would allow to account for natural spatial variation which can be particularly strong for some of the pollutants (e.g. NO$_2$) and should help increase the accuracy of the pollution estimates, hence reducing the measurement error variance. In addition such extension would allow to also evaluate chronic effects, which are generally dominated by spatial heterogeneity.  

%%\section{Conclusions}\label{Sec:Conclusions}

\section*{Acknowledgements}
We thank Margaret Douglass for the health data extraction. The work of the UK Small Area Health Statistics Unit (SAHSU) is funded by Public Health England as part of the MRC-PHE Centre for Environment and Health, funded also by the UK Medical Research Council. This paper does not necessarily reflect the views of Public Health England or the Department of Health. SAHSU has ethics and governance approvals to hold and use mortality data from the National Research Ethics Service - reference 7/LO/0846 and from the Health Research Authority Confidentially Advisory Group, reference HRA - 24/CAG/1039. We would also like to thank colleagues within the Environmental Research Group at King's College London and at the National Physical Laboratory for providing the pollutant measurements. MB and GF acknowledge partial support through the NERC/MRC funded project: NE/I00789X/1 ``Traffic Pollution and health in London''. MP and MB acknowledge partial support through the MRC funded project: MR/M025195/1 ``A general framework to adjust for missing confounders in observational studies''.

All authors declare no conflict of interests.

\bibliographystyle{unsrt}
\bibliography{Refs}

\begin{thebibliography}{10}

\bibitem{Xie:2014}
W~Xie, G~Li, D~Zao, X~Xie, Z~Wei, W~Wang, M~Wang, et~al.
\newblock Relationship between fine particulate air pollution and ischaemic
  heart disease morbidity and mortality.
\newblock {\em Heart}, 101:257--263, 2014.

\bibitem{Atkinson:2016}
R~Atkinson, A~Analitis, E~Samoli, G~Fuller, D~Green, I~Mudway, H~Anderson,
  et~al.
\newblock Short-term exposure to traffic-related air pollution and daily
  mortality in london, uk.
\newblock {\em J Expo Sci Environ Epidemiol}, 26:125--132, 2016.

\bibitem{Samoli:2016}
E~Samoli, R~Atkinson, A~Analitis, G~Fuller, D~Green, I~Mudway, H~Anderson,
  et~al.
\newblock Associations of short-term exposure to traffic-related air pollution
  with cardiovascular and respiratory hospital admissions in london, uk.
\newblock {\em Occup. Environ. Med}, 75:300--307, 2016.

\bibitem{Dai:2014}
L~Dai, A~Zanobetti, P~Koutrakis, and J~Schwartz.
\newblock Associations of fine particulate matter species with mortality in the
  united states: A multicity time-series analysis.
\newblock {\em Environ Health Perspect}, 122:837--842, 2014.

\bibitem{Pirani:2015}
M~Pirani, N~Best, M~Blangiardo, S~Liverani, R~Atkinson, and GW~Fuller.
\newblock Analysing the health effects of simultaneous exposure to physical and
  chemical properties of airborne particles.
\newblock {\em Environ Int}, 79:56--64, 2015.

\bibitem{Bobb:2015}
J~Bobb, L~Valeri, B~Henn, D~Christiani, R~Wright, M~Mazumdar, J~Godleski,
  et~al.
\newblock Bayesian kernel machine regression for estimating the health effects
  of multi-pollutant mixtures.
\newblock {\em Biostatistics}, 16:493--508, 2015.

\bibitem{Smith:2015}
G~Smith, Z~Bawa, Y~Macklin, R~Morbey, A~Dobney, S~Vordoulakis, and A~Elliot.
\newblock Using real-time syndromic surveillance systems to help explore the
  acute impact of the air pollution incident of march/april 2014 in england.
\newblock {\em Environ Res}, 136:500--504, 2015.

\bibitem{Finazzi:2013}
F~Finazzi, M~Scott, and A~Fass\'o.
\newblock A model-based framework for air quality indices and population risk
  evaluation, with an application to the analysis of scottish air quality data.
\newblock {\em J R Stat Soc Ser C}, 62:287--308, 2013.

\bibitem{Huang:2017}
G~Huang, M~Scott, and D~Lee.
\newblock Multivariate space-time modelling of multiple air pollutants and
  their health effects accounting for exposure uncertainty.
\newblock {\em Stat Med}, pages 1--15, 2017.

\bibitem{Atkinson:2010}
RW~Atkinson, GW~Fuller, RH~Anderson, RM~Harrison, and B~Armstrong.
\newblock Urban ambient particle metrics and health: a time-series analysis.
\newblock {\em Epidemiology}, 21:501--511, 2010.

\bibitem{Beddows:2015}
DC~Beddows, RM~Harrison, DC~Green, and GW~Fuller.
\newblock Receptor modelling of both particle composition and size distribution
  from a background site in london, {UK}.
\newblock {\em Atmospheric Chem Phys}, 15:10107--10125, 2015.

\bibitem{Font:2016}
A~Font and GW~Fuller.
\newblock Did policies to abate atmospheric emissions from traffic have a
  positive effect in {London}?
\newblock {\em Environ Pollut}, 218:463--474, 2016.

\bibitem{REVIHAPP}
{WHO Europe}.
\newblock {\em Review of evidence on health aspects of air pollution}.
\newblock WHO Regional Office for Europe, Copenhagen, Denmark, 2013.

\bibitem{Owens:2017}
EO~Owens, MM~Patel, E~Kirrane, TC~Long, J~Brown, I~Cote, MA~Ross, and others.
\newblock Framework for assessing causality of air pollution-related health
  effects for reviews of the national ambient air quality standards.
\newblock {\em Regul Toxicol Pharmacol.}, 88:332--337, 2017.

\bibitem{Mills:2016}
IC~Mills, RW~Atkinson, HR~Anderson, RL~Maynard, and DP~Strachan.
\newblock Distinguishing the associations between daily mortality and hospital
  admissions and nitrogen dioxide from those of particulate matter: a
  systematic review and meta-analysis.
\newblock {\em BMJ Open}, 6(7), 2016.

\bibitem{Shaddick:2002}
G~Shaddick and J~Wakefield.
\newblock Modelling daily multivariate pollutant data at multiple sites.
\newblock {\em J R Stat Soc Ser C}, 51:351--372, 2002.

\bibitem{Ruppert:2003}
D~Ruppert, MP~Wand, and RJ~Carroll.
\newblock {\em Semiparametric Regression}.
\newblock Cambridge University Press, 2003.

\bibitem{Crainiceanu:2005}
C~Crainiceanu, D~Rupert, and MP~Wand.
\newblock Bayesian analysis for penalized spline regression using winbugs.
\newblock {\em J Stat Softw}, 14:1--24, 2005.

\bibitem{bugs:2012}
D~Lunn, C~Jackson, N~Best, A~Thomas, and D~Spiegelhalter.
\newblock {\em The BUGS Book}.
\newblock Chapman \& Hall/CRC, 2012.

\bibitem{Dominici:2000}
F~Dominici, JM~Samet, and SL~Zeger.
\newblock Combining evidence on air pollution and daily mortality from the 20
  largest us cities: a hierarchical modelling strategy.
\newblock {\em J R Stat Soc Ser A}, 163:263--284, 2000.

\bibitem{Spiegelhalter:2002}
DJ~Spiegelhalter, NG~Best, BP~Carlin, and A~Van Der~Linde.
\newblock Bayesian measures of model complexity and fit.
\newblock {\em J R Stat Soc Ser B}, 64(4):583--639, 2002.

\bibitem{Peng:2006}
R~Peng, F~Dominici, and T~Louis.
\newblock Model choice in time series studies of air pollution and mortality.
\newblock {\em J R Stat Soc Ser A}, 169(2):179--203, 2006.

\bibitem{MacLehose:2007}
R~MacLehose, D~Dunson, A~Herring, and J~Hoppin.
\newblock Bayesian methods for highly correlated exposure data.
\newblock {\em Epidemiology}, 18:199--207, 2007.

\bibitem{Cefalu:2014}
M~Cefalu and F~Dominici.
\newblock Does exposure prediction bias health effect estimation? the
  relationship between confounding adjustment and exposure prediction.
\newblock {\em Epidemiology}, 25:583--590, 2014.

\bibitem{Williams:2014}
M~Williams, R~Atkinson, R~Anderson, and F~Kelly.
\newblock Associations between daily mortality in london and combined oxidant
  capacity, ozone and nitrogen dioxide.
\newblock {\em Air Qual Atmos Health}, 7:407--414, 2014.

\bibitem{Gamon:2014}
LF~Gamon, JM~White, and U~Wille.
\newblock Oxidative damage of aromatic dipeptides by the environmental oxidants
  no$_2^\ast$ and o$_3$.
\newblock {\em Org Biomol Chem}, 12:8280--8287, 2014.

\bibitem{Drechsler-Parks:1995}
DM~Drechsler-Parks.
\newblock Cardiac output effects of o$_3$ and no$_2$ exposure in healthy older
  adults.
\newblock {\em Toxicol Ind Health}, 11:99--109, 1995.

\bibitem{mauderly2009there}
Joe~L Mauderly and Jonathan~M Samet.
\newblock Is there evidence for synergy among air pollutants in causing health
  effects?
\newblock {\em Environ Health Perspect}, 117(1):1, 2009.

\end{thebibliography}
%\bibliography{Refs.bib}
%\bibliography{Refs}
\end{document}